\documentclass[12pt, english]{article}
\usepackage[utf8]{inputenc}
\usepackage{graphicx}

\usepackage{indentfirst} 
\usepackage{hyperref} 
\hypersetup{
            colorlinks=true,
            linkcolor=blue,
            urlcolor=violet,
            citecolor=violet,
            }

\usepackage{amsmath} 
\usepackage{amsfonts}

\usepackage{enumitem} 

\usepackage{float}

\usepackage{cite}

\usepackage{xcolor}

\usepackage{multirow}

\newcommand{\of}[1]{ \! \left( #1 \right ) }
\newcommand{\off}[1]{ \! \left [ #1 \right ] }

\newcommand{\re}{ \text{Re} }

\newcommand{\Z}{ \mathbb Z }
\newcommand{\C}{ \mathbb C }

\numberwithin{equation}{section}

\begin{document}
\begin{center}
\noindent{{\LARGE{Field response in the near-horizon limit of near-extremal 5-dimensional black holes}}}

\smallskip
\smallskip
\smallskip
\smallskip

\smallskip
\smallskip

\smallskip

\smallskip

\smallskip
\smallskip
\noindent{\large{Gaston Giribet$^{1}$, Juan Laurnagaray$^{2}$, Bryan Malpartida$^{2}$,}

\noindent{\large{Julio Oliva$^{3}$}, Osvaldo Santill\'an$^{4}$}}

\end{center}

\smallskip

\smallskip

\centerline{$^1$ Department of Physics, New York University, NYU}
\centerline{{\it 726 Broadway, New York, NY10003, USA.}}

\smallskip
\smallskip

\centerline{$^2$ Departamento de F\'{\i}sica, Universidad de Buenos Aires, UBA and IFIBA-CONICET}
\centerline{{\it Ciudad Universitaria, pabell\'on 1, 1428, Buenos Aires, Argentina.}}

\smallskip
\smallskip

\centerline{$^3$Departamento de F\'{\i}sica, Universidad de Concepci\'on}
\centerline{{\it Casilla, 160-C, Concepci\'on, Chile.}}

\smallskip
\smallskip

\centerline{$^4$Departamento de Matem\'{a}tica, Universidad de Buenos Aires, UBA and IMAS-CONICET}
\centerline{{\it Ciudad Universitaria, pabell\'on 1, 1428, Buenos Aires, Argentina.}}

\smallskip
\smallskip

\smallskip
\smallskip

\smallskip

\smallskip
\begin{abstract}
We study the scalar probe in the near-horizon region of near-extremal five-dimensional black holes and the problem of reattaching the asymptotic region. We consider the example of a Myers-Perry black hole with two independent angular momenta, for which the problem can be solved analytically in terms of the Riemann P-symbols and the confluent Heun special function. By prescribing leaking boundary conditions similar to those considered in the context of Kerr/CFT correspondence, we implement the attachment of the asymptotically flat region, matching the solutions in the near-horizon Myers-Perry geometry with those in the far region. This provides us with a set of explicit expressions for the field response in the background of five-dimensional stationary black holes near extremality, which enables us to highlight qualitative differences with the analogous problem in four dimensions.
\end{abstract}

\smallskip
\smallskip




\newpage

\section{Introduction}

The field dynamics in the high-redshift region of extremal and near-extremal Kerr black holes is governed by the infinite-dimensional local conformal symmetry group. The concrete realization of this idea is Kerr/CFT correspondence\cite{Guica:2008mu}, which states that extremal Kerr black holes admit a dual description in terms of a 2-dimensional conformal field theory (CFT$_2$). The near-horizon limit of extremal Kerr black holes is described by the so-called Near Horizon Extremal Kerr geometry (NHEK), which exhibits $SL(2, \mathbb R) \times U(1)$ symmetry\cite{Bardeen:1999px}. This geometry is closely related to squashed or stretched deformations of 3-dimensional Anti-de Sitter (AdS$_3$) space\cite{Bengtsson:2005zj,Anninos:2008fx}, and presents features that are reminiscent of the AdS$_2 \times S^2$ throat that emerges in near-horizon geometry of extremal Reissner-Nordström black holes. This suggests the possibility of a dual description of rapidly rotating black holes in terms of a quantum field theory. It was shown in Ref.\cite{Guica:2008mu} that, when the question is posed in terms of the asymptotic boundary conditions, then the exact $SL(2, \mathbb R) \times U(1)$ isometry of NHEK gets enhanced to the set of asymptotic symmetries generated by the two copies of Virasoro algebra, namely the symmetry algebra of a two-dimensional conformal field theory $\text{CFT}_2$. The conjecture that follows from this is that extremal Kerr black holes are dual to a $\text{CFT}_2$ with a central charge given by $c = 12 J$, with $J$ being the black hole angular momentum. Evidence supporting this statement comes from the observation that Cardy formula for the asymptotic growth of states in the $\text{CFT}_2$ precisely matches the Bekenstein-Hawking entropy of the black hole. Besides, the conjecture has been seen to work in a vast set of examples with remarkable success \cite{Lu:2008jk}; see \cite{Compere:2012jk} for a living review. Later, it was observed in \cite{Castro:2010fd} that infinite-dimensional conformal symmetry can also be found in non-extremally rotating black holes. This symmetry is manifested in the low frequency limit of field equations in Kerr background. Computationally, this is related to the fact that the field equations in such limit admit solutions in terms of hypergeometric equations, which transform nicely under $SL(2, \mathbb R)$. CFT observables such as reflection coefficients follow from the mix coefficients of the Kummer's functional relations between hypergeometric functions; therefore, what type of special functions appear in the computation of a given probe field in the near-horizon geometry is crucial for its dual interpretation. Among the interesting applications of Kerr/CFT, there are many that resort to the integrability of the field equations on the NHEK geometry in the probe approximation. In \cite{Porfyriadis:2014fja, Hadar:2014dpa}, for example, the conformal symmetry description of rotating black holes was employed to study the gravitational energy radiated by a massive probe star orbiting near-horizon zone of an extremal or near-extremal Kerr black hole. There, again, the crucial ingredient in the calculation is the conformal symmetry of the problem, which manifests itself in the fact that the solutions of the field equations admit to be expressed in terms of confluent hypergeometric functions, and so have a natural action of $SL(2, \mathbb R)$. This results in the bulk observables precisely reproducing the structure expected from the $\text{CFT}_2$ analysis. This is why solving analytically the field equation in the 5-dimensional case is of importance for a comparative analysis, cf. \cite{Bredberg:2009pv}. In this paper, we undertake a field theory computation similar to that of \cite{Porfyriadis:2014fja} in 5 spacetime dimensions, and show that, despite the apparent complexity of such a generalization, the field equations can still be solved analytically: We consider the extremal and near-extremal 5-dimensional black hole with two independent angular momenta in the near-horizon limit, which is described by the NHEK analogue for a rapidly rotating Myers-Perry black hole (hereafter referred to as NHEMP). We consider the solutions to the scalar field equation in this setting. As it happens in 4 dimensions, in the extremal case the field equations comprise confluent hypergeometric equations, whose solutions can be expressed in terms of Whittaker functions. In the near-extremal case, on the other hand, the field equations yield the Riemann-Papperitz differential equation, whose solutions, the so-called Riemann P-symbols, or Papperitz symbols, can also be expressed in terms of hypergeometric functions, with a consequent natural $SL(2,\mathbb{R})$ action. Finally, the spheroidal equation for the azimuthal angle reduces to the confluent Heun differential equation. Using all this, we explicitly compute the scalar field response in the near-horizon geometry, and then reattach the asymptotically flat region by considering leaking boundary conditions similar to those discussed in the context of Kerr/CFT. We discuss important differences between the 4- and 5-dimensional cases.

The paper is organized as follows: In section 2, we review the Myers-Perry black hole solution and its near-horizon geometry both for the extremal and near-extremal configurations. In section 3, we study the scalar field response on these geometries and show that the field equation in the near-horizon limit can be solved in terms of hypergeometric equations. This enables us to compute the response of the field excitations in the probe approximation, preserving certain boundary conditions on the horizon and in the asymptotic region.

\section{Myers-Perry and its near horizon limit}

\subsection{Myers-Perry solution}

The Myers-Perry (MP) solution\cite{Myers:1986un} is the 5-dimensional generalization of the Kerr solution, i.e. the metric of stationary black hole solution in asymptotically flat spacetime, with spherical horizon topology and two independent angular momenta. Its metric depends on three parameters, $a\in \mathbb{R}$, $b\in \mathbb{R}$ and $\mu\in \mathbb{R}_{\geq 0}$, which are related to the angular momenta and the mass; see (\ref{eq:MP_charges}) below. Written in Boyer-Lindquist type coordinates, its metric is
\begin{equation}
    \label{eq:mp_boyer_lindquist}
    \begin{aligned}
        ds^2 = & \, g_{\mu\nu}\, dx^{\mu}dx^{\nu} =  - d\tilde t^2 + \frac{\mu}{\tilde \rho^2} (d \tilde t - a \sin^2\tilde \theta d\tilde \phi - b \cos^2\tilde \theta d \tilde \psi)^2 \\
        & + \frac{\tilde r^2}{\Delta} \tilde \rho^2 d\tilde r^2 + \tilde \rho^2 d \tilde \theta^2 + 
        (\tilde r^2 + a^2) \sin^2 \tilde \theta d\tilde \phi^2 + (\tilde r^2 + b^2) \cos^2\tilde \theta d \tilde \psi^2 \, ,
    \end{aligned}
\end{equation}
with the functions
\begin{equation}
    \Delta = (\tilde r^2 + a^2)(\tilde r^2 + b^2) - \mu \tilde r^2, \qquad \tilde \rho^2 = \tilde r^2 + a^2 \cos^2\tilde \theta + b^2 \sin^2 \tilde \theta \, .
\end{equation}
Coordinates are $x^{\mu}=\{\tilde t, \tilde r,\tilde \phi, \tilde \psi , \tilde \theta \}$ for $\mu=0,1,2,3,4$, with ranges $\tilde t \in \mathbb{R}$, $\tilde r \in \mathbb{R}_{\geq 0}$, $\tilde \phi \in \off{0, 2\pi}$, $\tilde \psi \in \off{0, 2\pi}$, $\tilde \theta \in \off{0, \pi/2}$. The determinant of the metric is $  \det g = - \frac{1}{4} \tilde r^2 \, \tilde \rho^4 \sin^2{}(2 \tilde \theta) $.

The conserved charges associated to the Killing vectors $\partial_{\tilde t}$, $\partial_{\tilde \phi }$ and $\partial_{\tilde \psi }$ are given by
\begin{align}
	\label{eq:MP_charges}
    \mathcal M = \frac{3 \pi}{8 G} \mu, \ \ \, \mathcal J_{\tilde \phi } = \frac \pi{4 G} \mu a, \ \ \, \mathcal J_{\tilde \psi } = \frac \pi{4 G} \mu b,
\end{align}
respectively. These correspond to the Arnowitt-Deser-Misner mass and two angular momenta. Extremality condition for the MP solution corresponds to
\begin{align}
    \label{eq:MP_extremality_condition}
    \mu = (a + b)^2.
\end{align}
In this case, the degenerate event horizon is located at $\tilde r^2 = a b$.

For convenience, we can write the MP metric \eqref{eq:mp_boyer_lindquist} in terms of a new radial coordinate $\tilde u = \tilde r^2$. This coordinate will be useful later to study the near horizon limit. In terms of $\tilde u$, MP metric reads
\begin{align}
    \label{eq:MP_metric}
    ds^2 & = \of{\frac{\mu}{\rho^2} - 1} d\tilde t^2 + \rho^2 d\tilde \theta^2 - \frac{2 a \mu}{\rho^2} \sin^2\tilde \theta d\tilde t d\tilde \phi + 2 \frac{a b \mu}{\rho^2} \cos^2 \tilde \theta \sin^2 \tilde \theta d\tilde \phi d\tilde \psi - \frac{2 b \mu}{\rho^2} \cos^2 \tilde \theta d \tilde t d \tilde\psi \nonumber \\
    & + \of{\tilde u + a^2 + \frac{\mu a^2}{\rho^2} \sin^2\tilde \theta} \sin^2\tilde \theta d\tilde \phi^2 + \of{\tilde u + b^2 + \frac{\mu b^2}{\rho^2} \cos^2\tilde \theta} \cos^2\tilde \theta d\tilde \psi^2 + \frac{\rho^2}{4 \Delta} d\tilde u^2,
\end{align}
with
\begin{align}
	\label{eq:MP_auxiliary_functions}
    \Delta \equiv (a^2 + \tilde u)(b^2 + \tilde u) - \tilde u \mu, \ \ \ \rho^2 \equiv \tilde u + a^2 \cos^2\theta + b^2 \sin^2\theta \, ,
\end{align}
and
\begin{equation}
    \det g = -\frac{1}{16} \tilde \rho^4 \sin^2{}(2 \tilde \theta) \, .
\end{equation}

To follow the details of the calculation, it is also convenient to have at hand the components of the inverse metric; namely
\begin{equation}
    \begin{aligned}
        g^{\tilde \theta \tilde \theta} & = \frac{1}{\rho^2}, &   \qquad \qquad g^{\tilde u \tilde u} & = \frac{4 \Delta}{\rho^2} ,\\
        g^{\tilde t \tilde \phi} & = - \frac{a \, \mu \, (b^2 + \tilde u)}{\rho^2 \Delta} ,& g^{\tilde t \tilde \psi} & = - \frac{b \, \mu \, (a^2 + \tilde u)}{\rho^2 \Delta}, \\
        g^{\tilde t \tilde t} & = - 1 - \frac{\mu \, (a^2 + \tilde u) \, (b^2 + \tilde u)}{\rho^2 \Delta} ,& g^{\tilde \phi \tilde \psi} & = - \frac{a \, b \, \mu}{\rho^2 \Delta}, \\
        g^{\tilde \phi \tilde \phi} & = \frac{1}{a^2 + \tilde u}\off{\frac{1}{\sin^2\tilde \theta} - \frac{a^2 \, \mu \, (b^2 + \tilde u)}{\rho^2 \Delta}} ,& g^{\tilde \psi \tilde \psi} & = \frac{1}{b^2 + \tilde u}\off{\frac{1}{\cos^2\tilde \theta} - \frac{b^2 \, \mu \, (a^2 + \tilde u)}{\rho^2 \Delta}}.
    \end{aligned}\nonumber
\end{equation}

\subsection{Near-horizon limit of extremal Myers-Perry}

Now, let us study the geometry near the horizon in the extremal limit. Extremality condition is (\ref{eq:MP_extremality_condition}). The geometry of the near-horizon limit of the extremal MP black hole (NHEMP) is obtained as follows: first, define the rescaled time coordinate
\begin{align}{\label{coord.t}}
    t = \frac{2 \lambda}{\sqrt{ab}} \tilde t,
\end{align}
together with the shifted radial coordinate 
\begin{equation}
    \label{coord.u}
    u = \frac{\tilde u - a b}{\lambda (a + b)^2}.
\end{equation}
Then, boost the polar coordinates as follows
\begin{align}{\label{coord.angulos}}
    \phi = \tilde \phi - \frac{\tilde t}{a + b}, \ \ \, \psi = \tilde \psi - \frac{\tilde t}{a + b}, \ \ \, \theta = \tilde \theta \, .
\end{align}
Finally, NHEMP geometry is obtained by taking $\lambda \to 0$. This yields
\begin{align}
    d\hat s^2 & = \of{\frac{a b (a+b)^2}{4 \rho_0^2} u^2} dt^2
    + \rho_0^2 d\theta^2
    + \sqrt{a b} \of{(a + b) + \frac{a(a + b)^2}{\rho_0^2}} u \sin^2\theta dt d\phi \nonumber \\
    & + \sqrt{a b}\of{(a+b) + \frac{b(a + b)^2}{\rho_0^2}} u \cos^2\theta dt d\psi
    + \frac{a b(a + b)^2}{2 \rho_0^2} \sin^2\of{2 \theta} d\phi d\psi \\
    & + \frac{a (a+b)^2}{\rho_0^2} (a + b \sin^2\theta) \sin^2\theta d\phi^2 + \frac{b(a + b)^2}{\rho_0^2} (b + a \cos^2\theta) \cos^2\theta d\psi^2 + \frac{\rho_0^2}{4 u^2} du^2,\nonumber
\end{align}
with
\begin{align}\label{eq:rho_0}
    \rho_0^2 \equiv (a + b)(a \cos^2\theta + b \sin^2\theta);
\end{align}
see for instance \cite{Compere:2012jk}. The hat on the metric $d\hat s^2=\hat{g}_{\mu\nu }dx^{\mu } dx^{\nu }$ indicates that it corresponds to the metric of the near horizon geometry and it has to be distinguished from the MP metric $g_{\mu\nu}$. The components of the inverse of the NHEMP metric are
\begin{equation}
\begin{aligned}
    & \hat g^{tt} = -\frac 4{\rho_0^2 u^2}, \ \ \ \ \ \ \ \ \ \ \ \ \ \  \ \  \ & & \hat g^{uu} = 4 \frac{u^2}{\rho_0^2}, \\
    & \hat g^{t\phi} = 2 \sqrt{\frac ba} \frac 1{\rho_0^2} \frac 1u, & & \hat g^{\phi \phi} = \frac 1a\of{- \frac b{\rho_0^2} + \frac{b + a \cos^2\theta}{(a + b)^2} \frac 1{\sin^2\theta}}, \\
    & \hat g^{t \psi} = 2 \sqrt{\frac ab} \frac 1{\rho_0^2} \frac 1u, & & \hat g^{\psi \psi} = \frac 1b\of{- \frac a{\rho_0^2} + \frac{a + b \sin^2\theta}{(a + b)^2} \frac 1{\cos^2\theta}}, \\
    & \hat g^{\theta \theta} = \frac 1{\rho_0^2}, & & \hat g^{\phi \psi} = -\frac 1{(a+b)^2} - \frac 1{\rho_0^2},
\end{aligned}
\end{equation}
while the determinant of the metric is
\begin{align}
    \label{eq:NHEMP_det}
    \det \hat g = -\frac{a b(a+b)^4}{64} \rho_0^4 \sin^2\of{2 \theta}.
\end{align}

This geometry is the analog of the NHEK geometry in 4-dimensions, cf. \cite{Bardeen:1999px}.

\subsection{Near-Horizon limit of near-extremal Myers-Perry}

Now, let us consider the near-horizon geometry of the near-extremal MP metric (often referred to as near-NHEMP or NHnEMP). It describes the geometry near the horizon of a MP black hole whose mass is infinitesimally larger than the one needed to saturate the extremality bound \eqref{eq:MP_extremality_condition}. To approach the near-horizon limit we may use the same coordinate transformations as before; namely, we define coordinates as in (\ref{coord.t}), (\ref{coord.u}) and (\ref{coord.angulos}) together with a new parameter, $\eta $, that controls the departure from extremality; namely\footnote{The factor $\lambda^2$ multiplying $\eta$ is the minimal power for which all of the metric components become convergent when the limit $\lambda \to 0$ is taken.}
\begin{align}
    \mu = (a + b)^2 + \eta \lambda^2.
\end{align}

When equation \eqref{eq:MP_metric} is rewritten in terms of these new coordinates, and after taking the limit $\lambda \to 0$, one obtains the NHnEMP metric, whose components are
\begin{align}
    \begin{aligned}
    \label{eq:nNHEMP_metric}
    & \hat g_{tt} = \frac{a b(a + b)^2}{4 \rho_0^2} \off{u^2 + \frac{\rho_0^2}{(a+b)^6} \eta}, & & \hat g_{uu} = \frac{\rho_0^2}4 \frac{(a+b)^4}{(a+b)^4 u^2 - a b \eta}, \\
    & \hat g_{t \psi} = \frac{\sqrt{a b}(a+b)}{2 \rho_0^2} \off{\rho_0^2 + b(a +b)} u \cos^2\theta, & & \hat g_{t\phi} = \frac{\sqrt{a b}(a+b)}{2 \rho_0^2} \off{\rho_0^2 + a(a +b)} u \sin^2\theta, \\
    & \hat g_{\psi \psi} = \frac{b (a+b)^2}{\rho_0^2} \off{b + a\cos^2\theta} \cos^2\theta, & & \hat g_{\phi \phi} = \frac{a (a+b)^2}{\rho_0^2} \off{a + b\sin^2\theta} \sin^2\theta, \\
    & \hat g_{\theta \theta} = \rho_0^2, & & \hat g_{\phi \psi} = \frac{a b(a+b)^2}{4 \rho_0^2} \sin^2\of{2\theta},
    \end{aligned}
\end{align}
with
\begin{align}
    \rho_0^2 \equiv (a + b)(a \cos^2\theta + b \sin^2\theta).
\end{align}
One can easily verify that in the case $\eta = 0$ the components of the NHnEMP metric reduce to those of the NHEMP metric.

With the purpose of keep collecting useful formulae, let us write down the components of the inverse metric as well,
\begin{equation}
\begin{aligned}
    & \hat g^{tt} = -\frac 4{\rho_0^2} \frac{(a+b)^4}{(a+b)^4 u^2 - a b \eta}, & & \hat g^{uu} = \frac 4{\rho_0^2} \frac{(a+b)^4 u^2 - a b \eta}{(a+b)^4}, \\
    & \hat g^{t \psi} = 2 \sqrt{\frac ab} \frac 1{\rho_0^2} \frac{(a+b)^4 u}{(a+b)^4 u^2 - a b \eta}, & & \hat g^{t \phi} = 2 \sqrt{\frac ba} \frac 1{\rho_0^2} \frac{(a+b)^4 u}{(a+b)^4 u^2 - a b \eta}, \\
    & \hat g^{\theta \theta} = \frac 1{\rho_0^2}, & & \hat g^{\phi \psi} = -\frac 1{(a+b)^2} - \frac 1{\rho_0^2}\frac{(a+b)^4 u^2}{(a+b)^4 u^2 - a b \eta} 
\end{aligned}
\end{equation}
together with 
\begin{equation}
\begin{aligned}
    & \hat g^{\phi \phi} = -\frac 1{(a+b)^2} - \frac 1{\rho_0^2}\of{1 - \csc^2\theta + \frac{b^2 \eta}{(a+b)^4 u^2 - a b \eta}}, \\
    & \hat g^{\psi \psi} = -\frac 1{(a+b)^2} - \frac 1{\rho_0^2}\of{1 - \sec^2\theta + \frac{a^2 \eta}{(a+b)^4 u^2 - a b \eta}}.
\end{aligned}
\end{equation}
The determinant of the NHnEMP metric is independent of $\eta$, and so it coincides with \eqref{eq:NHEMP_det}.

\section{Scalar field response}

Having the explicit form of the NHEMP and  NHnEMP geometries, we are ready to study the field equation on these two spaces. We will study the wave equation for a massless\footnote{The solution for a massive scalar field follows straightforwardly with no major adaptation.} scalar field $\Phi$ in the probe approximation, and in different regimes of the MP geometry. That is to say, we will explicitly solve the field equation
\begin{align}
    \label{eq:wave_equation}
    \frac 1{\sqrt{-g}} \partial_\mu\of{\sqrt{-g} \, g^{\mu \nu} \partial_\nu} \Phi\of{t, u, \theta, \phi, \psi} = 0.
\end{align}
First, we will solve the problem in the NHEMP and NHnEMP geometries, and later we will proceed in a similar manner solving (\ref{eq:wave_equation}) in the far region of the full MP geometry. In all these cases the problem is separable, in the sense that the solution admits an ansatz of the form
\begin{align}
	\label{eq:wave_eq_separable_solution}
    \Phi\of{t, u, \theta, \phi, \psi} = R\of u \Theta\of \theta e^{i(-\omega t + k_1 \phi + k_2 \psi)},
\end{align}
with $\omega \in \mathbb{C}$ and $k_1,\, k_2 \in \mathbb{Z}$, and with $R\of u$ and $\Theta \of \theta$ being functions of the $u$ and $\theta$ coordinates, respectively.

\subsection{Fields in the near-horizon region of extremal black hole}

As said before, the wave equation in the NHEMP geometry is separable, as probably expected due to the integrability of the problem. This separation of variables leads to the angular and the radial ordinary differential equations for $\Theta\of \theta$ and $R\of u$. On the one hand, the equation for the angular coordinate takes the form
\begin{align}
	\label{eq:NHEMP_angular_equation}
    \frac{\partial_\theta(\sin\theta \cos\theta\, \partial_\theta \Theta\of \theta)}{\sin\theta \cos\theta} + \off{(k_1 + k_2)^2 \frac{a \cos^2\theta + b \sin^2\theta}{a+b} - \frac{k_1^2}{\sin^2\theta} - \frac{k_2^2}{\cos^2\theta}} \Theta\of \theta = -K_\ell\, \Theta\of \theta \, ,
\end{align}
where $K_\ell$ is the separation constant. On the other hand, the equation for the radial coordinate is
\begin{align}
	\label{eq:NHEMP_radial_equation}
    \partial_u\of{u^2 \partial_u R\of u} + \off{\frac A{u^2} + \frac Bu + \frac C4 } R\of u = \frac 14 K_\ell\, R\of u,
\end{align}
with
\begin{align}\label{eq:AByC}
    A = \omega^2, \ \ \ \ \, B =  \frac \omega{\sqrt{a b}} (a k_2 + b k_1), \ \ \ \ \, C = (k_1 + k_2)^2.
\end{align}

Separation constant $K_\ell$ is ultimately associated to the quantity that controls the asymptotic behaviour of $R\of u$. As in Kerr/CFT computations, the scaling exponent $\Delta$ can be read off from the large $u$ limit of the radial equation. That is to say, as in AdS/CFT computations, one proposes the asymptotic form $R\of u \simeq u^{- \Delta} + \ldots$ and inserts this into the equation above to find the condition
\begin{align}
	4 \Delta(\Delta - 1) + (k_1 + k_2)^2 - K_\ell = 0.
\end{align}
This gives two branches with different damping off conditions at large $r$; namely
\begin{align}
	\label{eq:delta_definition}
	\Delta_\pm = \frac 12 \pm \frac 12 \sqrt{1 + K_\ell - (k_1 + k_2)^2}.
\end{align}

Constant $K_{\ell }$ also enters in the solutions of the angular equation, of course. In equation \eqref{eq:NHEMP_angular_equation}, the $K_\ell$ can be thought of eigenvalues and $\ell $ represents the set of labels of the modes. These labels are specified by the eigenvalue problem of the 5-dimensional version of the spheroidal equation, cf. \cite{Porfyriadis:2014fja}. In fact, a more convenient notation for $\Theta\of \theta$ would include subindices $\ell$ labeling the solution of the corresponding $\ell$-mode; see (\ref{tumamatambien}) below.

One can easily check that Eq. \eqref{eq:NHEMP_angular_equation} is invariant under
\begin{align}
	a \leftrightarrow b, \ \ \ k_1 \leftrightarrow k_2, \ \ \ \theta \leftrightarrow \frac \pi2 - \theta \, 
\end{align}
as expected. After defining the variable $z = \sin^2\theta$, the angular equation takes the form
\begin{align}
	\label{eq:NHEMP_angular_equation_2}
	4 \partial_z(z(1-z)F_\ell\of z) + \off{(k_1 + k_2)^2 \frac{a(1-z) + b z}{a+b} - \frac{k_1^2}z - \frac{k_2^2}{1-z}} F_\ell\of z = -K_\ell F_\ell\of z,
\end{align}
with $F_\ell\of{\sin^2\theta} \equiv \Theta_\ell\of z$ and $z \in [0,1]$. This equation is of the Sturm-Liouville type. For such an eigenvalue problem, the corresponding eigenfunctions that obey certain boundary conditions yield an orthonormal basis. In our 5-dimensional case, the orthogonality relation for the elements of that basis reads
\begin{align}
	\int_0^{\frac{\pi}{2}} d\theta \sin \theta \cos \theta \, \Theta_\ell\of \theta \Theta_{\ell'}\of \theta = c_\ell \, \delta_{\ell, \ell'},\label{tumamatambien}
\end{align}
with $c_\ell$ being constants that can be reabsorbed in the normalization of the eigenfunctions. This orthogonality relation enables the decomposition of the scalar field in modes. By writing $F_\ell\of z$ as
\begin{align}
	F_\ell\of z = z^{k_1/2}(1-z)^{k_2/2} H_\ell\of z,
\end{align}
with $k_1$ and $k_2$ being the constants in \eqref{eq:wave_eq_separable_solution}, equation \eqref{eq:NHEMP_angular_equation_2} reduces to
\begin{align*}
	& \partial_z^2 H_\ell\of z + \off{\frac{(k_1+1)}{z} + \frac{(k_2+1)}{z-1}} \partial_z H_\ell\of z \\
	& \qquad + \frac 14 \off{(k_1 + k_2)^2 + 2 k_1 + 2 k_2 - \frac a{a+b}(k_1+k_2)^2 - \frac{b-a}{b+a} (k_1+k_2)^2 z - K_\ell} \frac{H_\ell\of z}{z (z-1)} = 0.
\end{align*}
This is the single confluent Heun differential equation
\begin{align}
	\label{eq:heun_diff_eq}
	\partial_z^2 H_\ell\of z + \of{\frac \gamma z + \frac \delta{z - 1}} \partial_z H_\ell\of z + \frac{\alpha z - q_\ell}{z(z-1)} H_\ell\of z = 0,
\end{align}
whose solution is the Heun special function $H\of{\alpha, q_\ell, \gamma, \delta; z}$ with parameters \cite{ronveaux1995heun}
\begin{equation}
\label{eq:heun_parameters}
\begin{aligned}
	& \alpha = -\frac{(k_1 + k_2)^2}4 \frac{b - a}{b + a}, & 4 & q_\ell = -(k_1 + k_2)^2 - 2k_1 - 2k_2 + (k_1 + k_2)^2 \frac a{a + b} + K_\ell, \\
	& \gamma = k_1 + 1, & & \delta = k_2 + 1.
\end{aligned}
\end{equation}
Therefore, the solution to the angular equation takes the form
\begin{align}
	\Theta_{\ell, k_1, k_2}\of \theta = (\sin\theta)^{k_1} (\cos\theta)^{k_2} H\of{\alpha,q_\ell, \gamma, \delta; \sin^2\theta},
\end{align}
with $H\of{\alpha, q_\ell, \gamma, \delta; z}$ being the solution to the confluent Heun equation evaluated in the parameters \eqref{eq:heun_parameters}. While it is not obvious from \eqref{eq:heun_diff_eq}-\eqref{eq:heun_parameters}, one can verify that the equation is still invariant under
\begin{align}
	a \leftrightarrow b, \ \ \ k_1 \leftrightarrow k_2, \ \ \ z \leftrightarrow 1 - z.
\end{align}

\begin{figure}[ht]
\begin{center}
\includegraphics[width=0.26\textwidth]{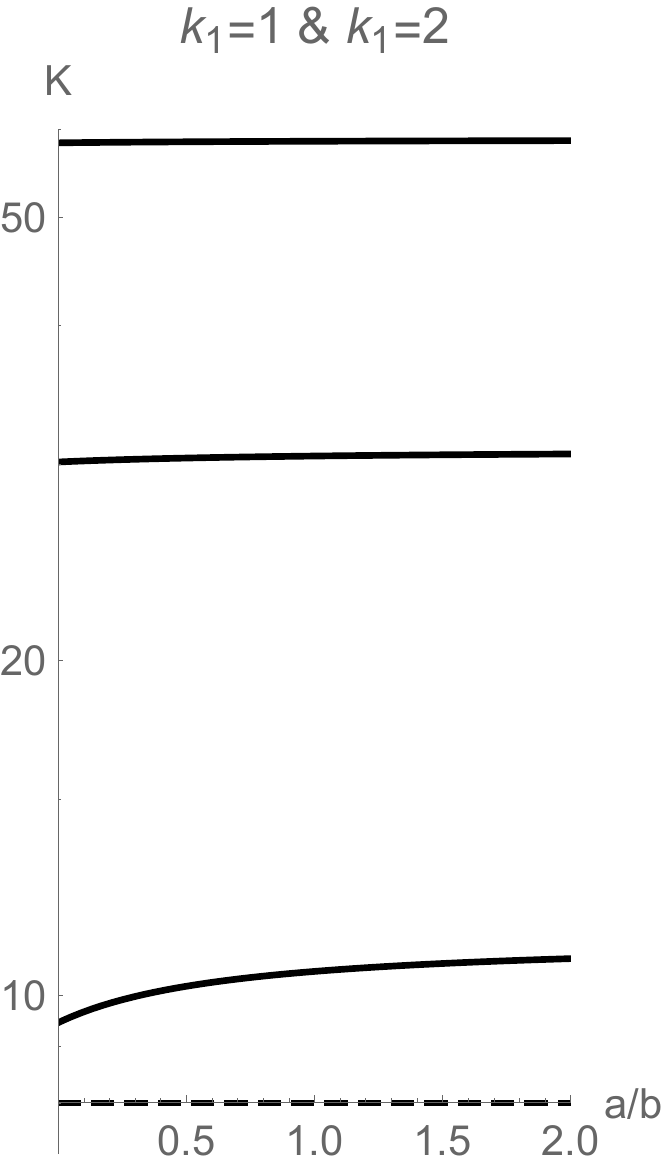}
\hspace{.01\textwidth} 
\includegraphics[width=0.26\textwidth]{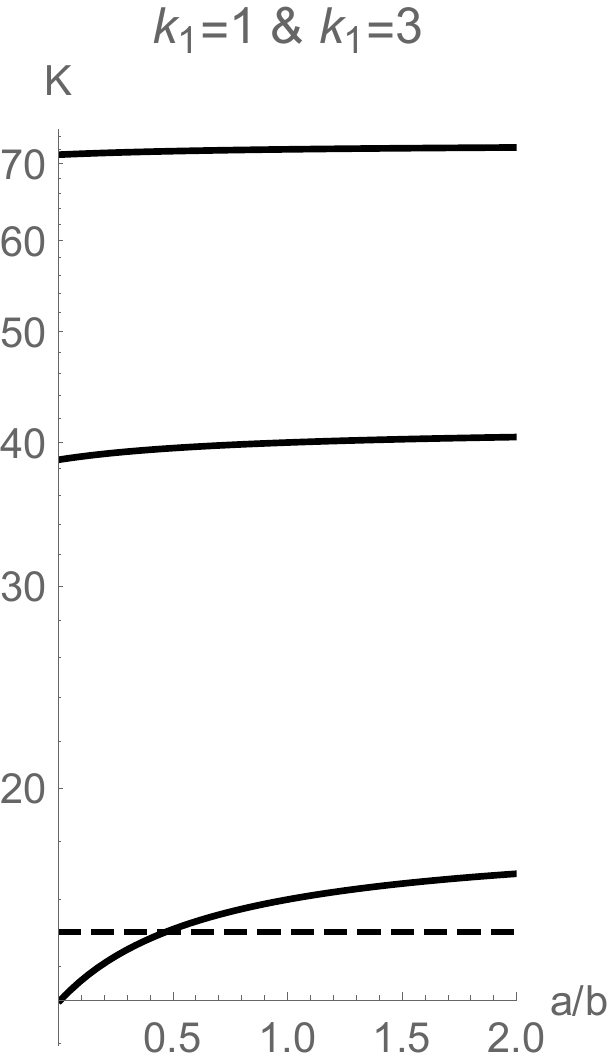}
\hspace{.01\textwidth} 
\includegraphics[width=0.26\textwidth]{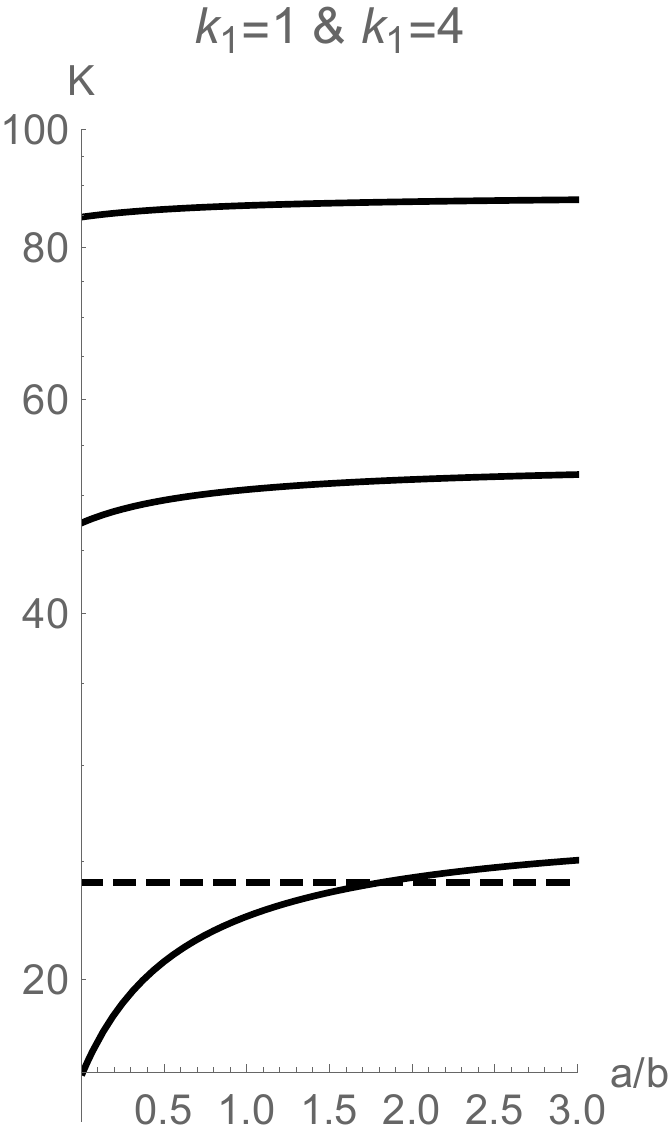}
\end{center}
\caption{\label{spheroidals}The solid lines represent the first three eigenvalues $K_{\ell}$ of the angular equation \eqref{eq:NHEMP_angular_equation}, for different values of $k_1$ and $k_2$, as a function of the ratio of the angular momenta $a/b$
. The dashed line represents a critical value of $K_{\ell}$ above which the eigenvalues $K_{\ell}$ lead to bound states in the near horizon geometry, instead of traveling waves; namely, for eigenvalues above the dashed line, the $\Delta_\pm$ in \eqref{eq:delta_definition} are real. The behavior is similar to that presented in the table with numerical values in page 9 of \cite{Bardeen:1999px}.}
\end{figure}

While function $H(a,b,c;z)$ is an extensively studied special function \cite{ronveaux1995heun} and it appears in many physics problems, including black holes in 3, 4 and 5 dimensions, it is certainly much more involved than other functions that are more familiar to us, such as hypergeometric functions. This is why we find illustrative to comment on some special cases for which the solution to the azimuthal equation (\ref{eq:heun_diff_eq}) notably simplifies. In fact, for specific values of the parameters, the confluent Heun function does become the hypergeometric function. This is useful because, among other things, it enables to connect the regular behavior at $z=0$ with that at $z=1$. One can immediately see that there is an $s$-wave mode for $k_1=k_2=K_{\ell}=0$, leading to a function $H_{\ell}(z)$ with a constant profile. A nontrivial case for which the solution to the angular equation simplifies as well is $k_{1}=-k_{2}\in \mathbb{Z}$; in that case, we get
\begin{equation}
K_n^{\text{(I)}}=4\left(  k_{2}+n\right)  \left(  1+k_{2}+n\right)  \text{\ \ with\ }%
n=0,1,2,...\ .
\end{equation}
It is worth noticing that these eigenvalues do not depend on the black hole angular momenta, in contrast to what happens in the generic case. A second family for which the eigenvalues can be found analytically is $a=b$, with arbitrary $k_{1,2}$. Also in this case, the angular momentum of the black hole does not appear in the angular equation, yielding
\begin{equation}
K_n^{\text{(II)}}=\frac{1}{2}\left(  k_{1}+k_{2}\right)  \left(  4+k_{1}+k_{2}\right)
+4\left(  1+k_{1}+k_{2}\right)  n+4n^{2}\text{\ \ with\ }n=0,1,2,...\ .
\end{equation}
When $k_{1}=-k_{2}$, the set $K_n^{\text{(II)}}$ reduces to $K_n^{\text{(I)}}$; nevertheless, the set $K_n^{\text{(I)}}$ is valid for
arbitrary values of the black hole angular momenta. For arbitrary values of $k_1,k_2\in \mathbb{Z}$ and $a,b\in \mathbb{R}$ one must solve the equation
numerically. Opposite to what happens in dimension 4, the equation for the angular dependence of the probe does depend on the angular momenta. Still, due to symmetry, to explore the parameter space it is sufficient to vary the quotient $0\leq a/b$, fixing $k_1$ and varying $k_2$; see Figure \ref{spheroidals}.

Now, let us focus on the radial equation \eqref{eq:NHEMP_radial_equation}; that is,
\begin{equation}
	\partial_u\of{u^2 \partial_u R\of u} + \of{\frac 14 (k_1 + k_2)^2 + \frac{\omega(a k_2 + b k_1)}{\sqrt{a b} \, u} + \frac{\omega^2}{u^2}} R\of u = \frac 14 K_\ell R\of u \, .
    \label{eq:radial_nhemp}
\end{equation}
After a change of variable, this equation becomes the confluent hypergeometric equation: defining $\zeta \equiv -2 i \omega u^{-1}$, it reads
\begin{equation}
	\label{eq:whittaker_diff_eq}
	\partial_\zeta^2 W\of \zeta + \of{-\frac 14 + \frac \lambda \zeta + \frac{1 - 4 \mu^2}{4 \zeta^2}}W\of \zeta = 0,
\end{equation}
with\footnote{Do not mistake the parameter $\mu$ here for the mass parameter in MP solution \eqref{eq:MP_charges}-\eqref{eq:MP_metric}. Since we are involved with extremal solutions, we will not longer need to refer to the black hole mass parameter.}
\begin{align}\label{lambda_y_mu}
	\lambda = i\frac{(b k_1 + a k_2)}{2 \sqrt{a b}}, \ \ \ \
	\mu^2 = \frac{K_\ell + 1 - (k_1 + k_2)^2}4.
\end{align}
The latter equation admits solutions of the form
\begin{align}
	\label{eq:whittaker_M_sol}
	W\of \zeta = \sum_{\epsilon = \pm 1} C_\epsilon M_{\lambda, \epsilon \mu}\of \zeta = \sum_{\epsilon = \pm 1} C_\epsilon\, \zeta^{\epsilon \mu + 1/2} e^{-\zeta/2} \, {}_1F_1\of{\frac 12 + \epsilon \mu -\lambda, 1 + 2 \epsilon \mu; \zeta},
\end{align}
where $C_\pm$ are arbitrary coefficients and where the confluent hypergeometric function ${}_1 F_1\of{\alpha, \gamma; \zeta}$ is defined as
\begin{align}
	{}_1 F_1\of{\alpha, \gamma; \zeta} = \sum_{s = 0}^\infty \frac{\Gamma\of \alpha \Gamma\of{\gamma + s}}{\Gamma\of \gamma \Gamma\of{\alpha + s}} \frac{\zeta^s}{s!}.
\end{align}

The set of solutions of the form \eqref{eq:whittaker_M_sol} is valid for $2 \mu \notin \Z$; it is not complete otherwise. In order to obtain a basis that is also valid for $2 \mu \in \Z$, one introduces the Whittaker functions
\begin{align}
	\label{eq:whittaker_W_sol}
	W_{\lambda, \mu}\of \zeta = \frac{\Gamma\of{-2\mu}}{\Gamma\of{1/2 - \mu -  \lambda}} M_{\lambda, \mu}\of \zeta
	+ \frac{\Gamma\of{2\mu}}{\Gamma\of{1/2 + \mu -\lambda}} M_{\lambda, -\mu}\of \zeta.
\end{align}

In our case, $\lambda = i(a k_2 + b k_1)/(2 \sqrt{a b})$, $\mu = \sqrt{1 + K_\ell - (k_1 + k_2)^2}/2$, and $\zeta = -2 i \omega / u$. Also, from \eqref{eq:delta_definition} we have $\Delta_+ = 1/2 + \mu$. Therefore, we can write the solution in terms of the scaling dimensions $\Delta_\pm$ and the parameter
\begin{align}
    \label{eq:whittaker_coeff_ilambda}
	p_{1,2} \equiv \frac 12 \sqrt{\frac ab} k_2 + \frac 12 \sqrt{\frac ba} k_1 = -i\lambda\, .
\end{align}

Functions $M_{\lambda, \pm \mu} \of \zeta$ and $W_{\lambda, \pm \mu}\of \zeta$ obey some functional relations that it might be convenient to collect as they will be useful for our analysis. In particular, from \eqref{eq:whittaker_W_sol} and the series expansion \eqref{eq:whittaker_M_sol} we obtain the following asymptotic expansion, valid for small $|u|$ (i.e. large $|\zeta|$),
\begin{equation}
M_{ip, \Delta_+ - \frac 12}\of{-\frac{2 i \omega}u} \simeq \frac{\Gamma\of{2 \Delta_+}}{\Gamma\of{\Delta_+ - ip}} e^{-\frac{i \omega}u}\of{-\frac{2 i \omega}u}^{-ip}  + \frac{\Gamma\of{2 \Delta_+}}{\Gamma\of{\Delta_+ + ip}} \of{-\frac{2 i \omega}u}^{ip} e^{\frac{i \omega}u \pm (\Delta_+ - ip) i \pi}
\end{equation}
with 
\begin{equation}
\Delta_+ - \frac 12 \mp i p \neq - \frac 12, - \frac 32, \ldots \,
\end{equation}
and
\begin{align}
	W_{ip, \Delta_+ - 1/2}\of{-\frac{2 i \omega}u} \simeq e^{\frac{i \omega}u} \of{-\frac{2 i \omega}u}^{ip}.
\end{align}
For large $|u|$ (i.e. small $|\zeta|$), the asymptotic behaviour is
\begin{align}
	M_{ip, \Delta_+ - \frac 12}\of{-\frac{2 i \omega}u} \simeq \of{-\frac{2 i \omega}u}^{\Delta_+}, \qquad
	2 \Delta_+ - 1 \neq -1, -2, -3, \ldots
\end{align}
and either
\begin{align}
    \label{eq:whittaker_W_sol_near_horizon_big_delta}
	W_{ip, \Delta_+ - \frac 12}\of{-\frac{2 i \omega}u} \simeq \frac{\Gamma\of{2 \Delta_+ - 1}}{\Gamma\of{\Delta_+ - ip}} \of{-\frac{2 i \omega}u}^{1 - \Delta_+}, \qquad \re\off{\Delta_+} \geq 1,\, \Delta_+ \neq 1
\end{align}
or
\begin{align}
    \label{eq:whittaker_W_sol_near_horizon}
	& W_{ip, \Delta_+ - \frac 12}\of{-\frac{2 i \omega}u} \simeq \frac{\Gamma\of{2 \Delta_+ - 1}}{\Gamma\of{\Delta_+ - ip}} \of{-\frac{2 i \omega}u}^{1 - \Delta_+} \nonumber \\
	& \qquad \qquad + \frac{\Gamma\of{1 - 2 \Delta_+}}{\Gamma\of{1 - \Delta_+ - ip}} \of{-\frac{2 i \omega}u}^{\Delta_+}, \qquad \frac 12 \leq \re\off{\Delta_+} < 1,\, \Delta_+ \neq \frac 12 \, .
\end{align}

All these formulae will be useful when solving the matching problem between the solution in the near horizon geometry and that in the flat region.

\subsection{Fields in the near-horizon region of near-extremal black hole}

Now, let us perform a similar analysis for the near-extremal case ($\eta \neq 0$). In this case, the near-horizon analysis is much more involved; however, it can still be solved analytically by following the same method: first, in equation \eqref{eq:wave_equation} with the background \eqref{eq:nNHEMP_metric} we propose a separable ansatz, yielding
\begin{align}
    \label{eq:nNHEMP_radial_equation}
    & \partial_u\off{\of{u^2 - \frac{a b \eta}{(a+b)^4}}\partial_uR\of u} + \bigg[ \frac{(k_1+k_2)^2 - K_\ell}{4} \nonumber \\
    & \qquad + \frac{(a + b)^4}{(a + b)^4 u^2 - a b \eta} \of{\omega^2 + \frac{a k_2 + b k_1}{\sqrt{ab}} \omega u + \frac{(a k_2 + b k_1)^2}{4 (a + b)^4} \eta} \bigg] R\of u = 0
\end{align}
which in the limit $\eta \to 0$ reduces to that of the NHEMP case, cf. \eqref{eq:NHEMP_radial_equation}. Then, we notice that the differential equation \eqref{eq:nNHEMP_radial_equation} is of the class
\begin{align}
    \partial_u\of{(u - \alpha)(u - \beta)\partial_u R\of u} + \off{\frac{Au + B}{(u - \alpha)(u - \beta)} + C} R\of u = 0,
\end{align}
with its constants being
\begin{align}
    A =  \frac{b k_1 + a k_2}{\sqrt{ab}} \omega, \qquad B = \omega^2 + \frac{(a k_2 + b k_1)^2}{4(a+b)^4} \eta, \qquad C = \frac{(k_1 + k_2)^2 - K_\ell}{4},
\end{align}
and its roots\footnote{This procedure is valid for complex roots $\alpha, \beta \in \C$.}
\begin{align}
    \alpha = - \beta = \frac{\sqrt{a b \eta}}{(a + b)^2}.
\end{align}
Then, we rewrite
\begin{align}
    \frac{Au + B}{(u - \alpha)(u - \beta)} = \frac {D}{u - \alpha} + \frac {E}{u - \beta},
\end{align}
with
\begin{align}
    D = \frac{A \alpha + B}{\alpha - \beta}, \qquad E = \frac{A \alpha - B}{\alpha - \beta},
\end{align}
and get
\begin{align}
    \partial_u\off{(u-\alpha)(u-\beta)\partial_u R\of u} + \of{\frac {D}{u - \alpha} + \frac {E}{u - \beta} + C} R\of u = 0.
\end{align}
Defining $u = 1/t$, the equation takes the form
\begin{equation}
    \label{eq:nNHEMP_radial_equation_2}
    \begin{aligned}
        \partial_t^2T\of t & + \of{\frac 1{t - t_1} + \frac 1{t - t_2}} \partial_t T\of t \\
        & + \off{\frac C{\alpha \beta\, t} - \frac{D}{\alpha^2 \beta (t - t_1)} - \frac{E}{\alpha \beta^2 (t - t_2)}} \frac{T\of t}{t(t-t_1)(t-t_2)} = 0,
    \end{aligned}
\end{equation}
where $T\of t = R\of{1/t}$ and
\begin{align}
    t_1 = \frac 1\alpha, \qquad t_2 = \frac 1\beta  = - \frac 1\alpha\, .
\end{align}
Finally, we can compare \eqref{eq:nNHEMP_radial_equation_2} with the so-called Riemann-Papperitz differential equation\cite{abramowitz1970handbook}, whose generic expression is given by
\begin{align}
    \label{eq:riemann_diff_eq}
    0 & = \partial_t^2 T\of t + \of{\frac{1-a_1-a_2}{t-t_0} + \frac{1-b_1-b_2}{t-t_1} + \frac{1-c_1-c_2}{t-t_2}} \partial_t T\of t \, + \Big[\frac{a_1 a_2(t_0 - t_1)(t_0 - t_2)}{t - t_0}\nonumber \\
    &  \ \ + \frac{b_1 b_2(t_1 - t_0)(t_1 - t_2)}{t - t_1} + \frac{c_1 c_2(t_2 - t_0)(t_2 - t_1)}{t - t_2}\Big] \frac{T\of t}{(t - t_0)(t - t_1)(t - t_2)}.
\end{align}
The regular singular points of this equation are $t_0$, $t_1$ and $t_2$, with the pairs of exponents for each point being $a_1$, $a_2$; $b_1$, $b_2$; $c_1$, $c_2$, respectively. These exponents are constrained to fulfill the condition
\begin{align}
    \label{eq:riemann_diff_eq_parameter_constraint}
    a_1 + a_2 + b_1 + b_2 + c_1 + c_2 = 1,
\end{align}
which is equivalent to demanding the sum of the numerators of the terms multipliying $\partial_t T\of t$ to equal 2. Condition \eqref{eq:riemann_diff_eq_parameter_constraint} is actually satisfied for \eqref{eq:nNHEMP_radial_equation_2}. In fact, by comparing \eqref{eq:nNHEMP_radial_equation_2} and \eqref{eq:riemann_diff_eq}, we get
\begin{align}
    a_1 + a_2 = 1, \qquad b_1 + b_2 = 0, \qquad c_1 + c_2 = 0.
\end{align}
In order to completely identify the parameters in \eqref{eq:nNHEMP_radial_equation_2} with those in \eqref{eq:riemann_diff_eq}, we have to solve the system
\begin{align}
    a_1 a_2 (t_0 - t_1)(t_0 - t_2) & = \frac{A}{\alpha \beta}, \\
    b_1 b_2 (t_1 - t_0)(t_1 - t_2) & = - \frac{D}{\alpha^2 \beta}, \\
    c_1 c_2 (t_2 - t_0)(t_2 - t_1) & = - \frac{E}{\alpha \beta^2},
\end{align}
which in our case, since $t_0 = 0$ and $t_2 = - t_1$, reduces to
\begin{align}
    a_1 a_2 t_1^2 = \frac A{\alpha^2}, \qquad 2 b_1 b_2 t_1^2 = \frac{D}{\alpha^3}, \qquad 2 c_1 c_2 t_1^2 = - \frac{E}{\alpha^3}.
\end{align}
This yields six algebraic equations, which are solved by
\begin{align}
    & a_1 = \frac{1 - \sqrt{1 - 4 C}}2, & & a_2 = \frac{1 + \sqrt{1 - 4 C}}2, \\
    & b_1 = \frac 1{2 \alpha} \sqrt{-A \alpha - B } & & b_2 = - b_1, \\
    & c_1 = \frac 1{2 \alpha} \sqrt{A \alpha - B} & & c_2 = - c_1.
\end{align}
Notice also that
\begin{equation}
    A \alpha \pm B  = \pm \of{-\omega \mp \frac{a k_2 + b k_1}{2 (a + b)^2} \sqrt \eta}^2 = \pm \of{-\omega \mp p \alpha}
\end{equation}
where $p=p_{1,2}$ is given in \eqref{eq:whittaker_coeff_ilambda}. In summary, the radial equation \eqref{eq:nNHEMP_radial_equation_2} is a Riemann-Papperitz equation\footnote{Do not mistake the parameter $z$ here for the variable introduced in \eqref{eq:NHEMP_angular_equation_2}.} whose exponents are given by
\begin{align}
    \label{eq:riemann_diff_eq_parameters}
    \begin{aligned}
    & a_1 = \frac 12 + \frac 12 \sqrt{1 + K_\ell - (k_1 + k_2)^2} = \Delta_+, & & a_2 = \frac 12 - \frac 12 \sqrt{1 + K_\ell - (k_1 + k_2)^2} = \Delta_-, \\
    & b_1 = \frac 1{2 \alpha}\sqrt{-(-\omega - p \alpha)^2} = -i \frac \omega{2 \alpha} - i \frac{p}{2}, & & b_2 = - b_1, \\
    & c_1 = \frac 1{2 \alpha} \sqrt{-(-\omega + p \alpha)^2} = -i \frac{\omega}{2\alpha} + i \frac{p}{2}, & & c_2 = - c_1,
    \end{aligned}
\end{align}
and
\begin{align}
    t_0 = 0,
    \qquad t_1 = \frac 1\alpha = \frac{(a+b)^2}{\sqrt{a b \eta}},
    \qquad t_2 = -\frac 1\alpha = - \frac{(a+b)^2}{\sqrt{a b \eta}}.
\end{align}

Interestingly enough, the solution of the Riemann-Papperitz equation \eqref{eq:riemann_diff_eq} admits to be written in terms of hypergeometric functions as well. In fact, one such solution is given by (see $u_5$ in page 284 of \cite{whittaker_watson_1996})
\begin{align}
    T\of t = \of{\frac{t - t_1}{t - t_2}}^{b_1} \of{\frac t{t - t_2}}^{a_1} {}_2 F_1 \of{\ell, m, s, \xi}, \qquad \xi = \frac{(t - t_1) t_2}{(t - t_2) t_1},
\end{align}
with
\begin{align}
    \ell = a_1 + b_1 + c_1, \ \ \ m = a_1 + b_1 + c_2 = a_1 + b_1 - c_1, \ \ \ s = 1 + b_1 - b_2 = 1 + 2 b_1.
\end{align}

A more general solution to \eqref{eq:riemann_diff_eq} is given by the linear combination
\begin{align}
    \label{eq:riemann_diff_eq_general_solution}
    T\of t = F\of \xi = (1 - \xi)^{a_1} \xi^{b_1}\of{C_1 h\of \xi + C_2 k\of \xi},
\end{align}
with (see $y_5$ ibid.)
\begin{align}
    \label{sec3.3:h_sol}
    h\of \xi = {}_2 F_1 \of{\ell, m; s; \xi},
\end{align}
and (see $y_6$ ibid.)
\begin{align}
    \label{eq:riemann_diff_eq_h_solution}
    k\of \xi = \xi^{1 - s} {}_2 F_1\of{\ell - s + 1, m - s + 1; 2 - s; \xi},
\end{align}
with $C_1$ and $C_2$ being two arbitrary constants.

The singular point $t = t_1$ corresponds to $\xi = 0$; the solutions expanded around this point are valid near the horizon. The point $t = 0$, on the other hand, corresponds to $\xi = 1$, and the solutions expanded around this point are valid in the asymptotic region $u \to \infty$. Let us start by analyzing the solutions close to the horizon, i.e. around $\xi = 0$. We observe that the second term in \eqref{eq:riemann_diff_eq_general_solution} goes like $\xi^{- b_1}$ and then it violates the incoming condition at the horizon. Then, we have to set $C_2 = 0$, so that the solution takes the form
\begin{align}
    F\of \xi = C_1 (1 - \xi)^{a_1} \xi^{b_1}\, {}_2 F_1 \of{\ell, m; s; \xi}.
\end{align}
In order to see what happens at infinity, we resort to the Kummer relations and write
\begin{align}
    F\of \xi & = C_1 (1-\xi)^{a_1} \xi^{b_1} \frac{\Gamma\of s \Gamma\of{s - \ell - m}}{\Gamma\of{s - \ell} \Gamma\of{s - \ell}}\, {}_2 F_1\of{\ell, m; \ell + m - s + 1; 1 - \xi} + C_1 \of{1 - \xi}^{a_1} \xi^{b_1}\nonumber \\
    & \qquad  \of{1 - \xi}^{s - \ell - m} \frac{\Gamma\of s \Gamma\of{\ell + m - s}}{\Gamma\of{\ell} \Gamma\of{\ell}} {}_2 F_1\of{s - \ell,s - m; s - \ell - m + 1; 1 - \xi}.
\end{align}
keeping in mind that
\begin{align}
    & a_1 = \Delta_+, \qquad \ell = \Delta_+ - \frac{i \omega}{\alpha}, \qquad m = \Delta_+ - i p, \qquad s = 1 + i\of{-\frac{\omega}{\alpha} - p}, \\
    & \ell + m - s = 2 \Delta_+ - 1, \qquad a_1 + s - \ell - m = 1 - \Delta_+.
\end{align}
Then, close to $\xi \simeq 1$, we find that the solution behaves as follows
\begin{align}
    F\of \xi & \simeq C_1 (1-\xi)^{\Delta_+} \frac{\Gamma\of{1 - \frac{i\omega}{\alpha} - i p} \Gamma\of{1 - 2 \Delta_+}}{\Gamma\of{1 - \Delta_+ - i p} \Gamma\of{1 - \Delta_+ - \frac{i \omega}{\alpha}}} \nonumber \\
    & \qquad + C_1 \of{1 - \xi}^{1 - \Delta_+} \frac{\Gamma\of{1 - \frac{i \omega}{\alpha} - i p} \Gamma\of{2 \Delta_+ - 1}}{\Gamma\of{\Delta_+ - \frac{i \omega}{\alpha}} \Gamma\of{\Delta_+ - i p}}.
\end{align}

These behaviors are useful to solve the matching condition when reattaching the asymptotic region.

\subsection{Reattaching the asymptotic region}

Now, we move to study the scalar field equation \eqref{eq:wave_equation} formulated on the full MP geometry and then analyze the behavior in the far region. We will study the solution in the extremal case and for large $\tilde u$, and then we will match this solution with the solution we obtained for the problem in the near horizon zone. The type of matching condition we will consider are the same leaking boundary conditions studied in \cite{Porfyriadis:2014fja} in the context of Kerr/CFT in 4 dimensions. 

Considering in \eqref{eq:wave_equation} the ansatz 
\begin{equation}
\Phi(\tilde t, \tilde r, \tilde \theta, \tilde \phi, \tilde \psi) = U(\tilde r) \Theta(\tilde \theta) e^{i(-\tilde \omega \tilde t + k_1 \tilde\phi + k_2 \tilde\psi)}
\end{equation}
we get
\begin{equation}
    \begin{aligned}\label{eq:MP_wave_equation}
        0 & = \frac{1}{\tilde r \, U\of{\tilde r}} \partial_{\tilde r}\of{\frac{1}{\tilde r} \Delta \partial_{\tilde r} U\of{\tilde r}} + \frac{1}{\Delta} \of{\tilde \alpha + \tilde \beta \, \tilde r^2 + \tilde\gamma \, \tilde r^4} - \frac{k_1^2}{\sin^2\tilde \theta} - \frac{k_2^2}{\cos^2\tilde \theta}+\\
        & \qquad  \tilde \omega^2 \tilde \rho^2  + \frac{1}{\cos\tilde \theta \sin\tilde\theta} \frac{1}{\Theta(\tilde \theta)} \partial_{\tilde \theta} \of{\cos\tilde\theta \sin\tilde\theta \partial_{\tilde \theta} \Theta(\tilde\theta)} \, ,
    \end{aligned}
\end{equation}
with
\begin{equation}
    \begin{aligned}
        \tilde \alpha & = -(a^2 - b^2)(a^2 k_2^2 - b^2 k_1^2) + (a k_2 + b k_1)^2 \, \mu - 2 a b (a k_2 + b k_1) \mu \tilde \omega + a^2 b^2 \, \mu \, \tilde \omega^2 \, ,\\
        \tilde \beta & = (a^2 - b^2)(k_1^2 - k_2^2) - 2(a k_1 + b k_2) \, \mu \, \tilde \omega + (a^2 + b^2) \, \mu \, \tilde \omega^2 \, ,\\
        \tilde \gamma & = \mu \, \tilde \omega^2 \, .
    \end{aligned}
\end{equation}
Defining $\tilde u = \tilde r^2$, the equation for the radial and azimuthal coordinates reads
\begin{equation}
    \begin{aligned}\label{eq:MP_wave_equation}
        0 & = \frac{4}{\tilde R\of{\tilde u}} \partial_{\tilde u}\of{\Delta \partial_{\tilde u} \tilde R\of{\tilde u}} + \frac{1}{\Delta} \of{\tilde \alpha + \tilde \beta \, \tilde u + \tilde\gamma \, \tilde u^2}\, + \\
        & \qquad  \tilde \omega^2 \tilde \rho^2 - \frac{k_1^2}{\sin^2\tilde \theta} - \frac{k_2^2}{\cos^2\tilde \theta} + \frac{1}{\Theta(\tilde \theta)} \frac{\partial_{\tilde \theta} \of{\cos\tilde\theta \sin\tilde\theta \partial_{\tilde \theta} \Theta(\tilde\theta)}}{\cos\tilde \theta \sin\tilde\theta} \, ,
    \end{aligned}
\end{equation}
for $\tilde R\of{\tilde u} = R\of u$; see coordinates change in (\ref{coord.u}). Finally, 
we find that the equation separates in its radial and angular parts; namely
\begin{equation}
    \label{eq:radial_u_coordinate}
    \frac{4}{R\of{u}} \partial_{u}\of{u^2\partial_{u} R\of{u}} + \frac{1}{\lambda^2 \mu^2 u^2} \of{\tilde \alpha + \tilde \beta \, (\lambda \mu u+a b) + \tilde\gamma \, (\lambda \mu u+ab)^2}+\tilde\omega^2\lambda\mu u=K_{\ell}
\end{equation}
and
\begin{equation}{\label{eq:angular_theta_coordinate}}
\tilde \omega^2 (a+b)(a\cos^2{\theta}+b\sin^2{\theta}) - \frac{k_1^2}{\sin^2 \theta} - \frac{k_2^2}{\cos^2 \theta} + \frac{1}{\Theta(\theta)} \frac{\partial_{\theta} \of{\cos\theta \sin\theta \partial_{\theta} \Theta(\theta)}}{\cos\theta \sin\theta}=-K_{\ell} \, ,
\end{equation}
respectively. More succinctly, we have 
\begin{equation}
\label{eq:mp:radial_eq}
\begin{aligned}
        \partial_{u}&\of{u^2 \partial_{u}R\of{u}}+
        \frac{1}{4}\off{\frac{\alpha(\tilde\omega)}{u^2}+\frac{\beta(\tilde\omega)}{u}+\gamma(\tilde\omega)- K_\ell+ u \lambda \, \mu \tilde\omega^2} R(u) = 0
\end{aligned}
\end{equation}
with
\begin{equation}\label{eq:coeficientes}
        \alpha = \frac{\tilde \alpha + \tilde \beta a b + \tilde \gamma a^2 b^2}{\lambda^2\mu^2}\, , \ \ \ \
        \beta  = \frac{\tilde \beta +2\tilde\gamma a b}{\lambda\mu}\, , \ \ \ \ 
        \gamma = \tilde\gamma.
    \end{equation}

We can read the relation between the NHEMP frequency $\omega$ and MP frequency $\tilde \omega$ by comparing the exponents in the dependence on $t$, $\phi$, $ \psi$ and $\tilde t$, $\tilde \phi$, $ \tilde \psi$. This yields
\begin{equation}
    {(-\tilde\omega\tilde t+k_1\tilde\phi+k_2\tilde\psi)}={{\frac{\sqrt{a b}}{2\lambda}\of{-\tilde\omega+\frac{k_1+k_2}{a+b}}t +k_1\phi+k_2\psi}}={ {-\omega t+k_1\phi+k_2\psi}} \, ;
\end{equation}
so that
\begin{equation}
    \label{eq:omega-tilde}
    2 \lambda \omega = -\frac{\sqrt{a b}}{a + b}\of{k_1 + k_2} + \tilde \omega \sqrt{a \, b} \, .
\end{equation}
In the $\lambda \to 0$ limit, we see the frequencies $\omega$ on the NHEMP geometry to flow to the unique frequency $\tilde{\omega }=(k_1+k_2)^2/(a+b)$ in the asymptotic region. The latter is analogous to the threshold frequency in superradiance.

We can use the relation (\ref{eq:omega-tilde}) between frequencies to implement the matching condition. In order to do so, we need to match the small $u$ behavior of the solution in the far region, with the large $u$ behavior of the solution in the near horizon region. This would build a bridge between both regimes. In the $u\ll 1$ limit, the radial equation becomes
\begin{equation}
\begin{aligned}
    \partial_{u}&\of{u^2 \partial_{u}R\of{u}}+
    \off{\frac{\alpha}{4 u^2}+\frac{\beta}{4 u}+\frac{\gamma}{4}- \frac{K_\ell}{4}} R(u) = 0.
\end{aligned}
\end{equation}
This permits to verify that equations \eqref{eq:radial_u_coordinate} and \eqref{eq:angular_theta_coordinate}, once the frequency \eqref{eq:omega-tilde} is replaced in the coefficients \eqref{eq:coeficientes}, coincide with the NHEMP analogs \eqref{eq:NHEMP_radial_equation} and \eqref{eq:NHEMP_angular_equation} in the limit $\lambda\to 0$. It is also worth noticing that considering \eqref{eq:omega-tilde} and taking the limit $\lambda \to 0$, i.e. $\tilde \omega =  (k_1 + k_2)/(a + b)$ in \eqref{eq:mp:radial_eq}, the coefficients $\alpha$ and $\beta$ identically vanish. As a consistency check, in the limit $\lambda\to 0$ we get $\alpha=A$, $\beta=B$ and $\gamma=C$, as defined in \eqref{eq:AByC}.

Now, taking the opposite limit, namely $u\gg 1$ large, in the extremal case $\mu = (a + b)^2$ the equation \eqref{eq:mp:radial_eq} becomes
\begin{equation}
    \partial_{u} \off{u^2  \partial_{u}R\of{u}} + \off{\frac{\Omega^2 \, u }{4} + \frac{1}{4}-q^2} R(u) = 0 \, ,
\end{equation}
with
\begin{equation}
    q^2 \equiv \frac{1+K_l - \tilde \omega^2(a+b)^2}{4}, \qquad \Omega^2 \equiv (a+b)^2 \tilde \omega^2 \, .
\end{equation}
The condition given in \eqref{eq:omega-tilde} with $\lambda=0$ yields
\begin{equation}
    q^2 = \frac{1+K_l - (k_1+k_2)^2}{4}, \qquad \Omega^2 = (k_1+k_2)^2 \qquad \tilde\omega = \frac{k_1+k_2}{a+b}\, .
\end{equation}

It is remarkable that the solution to this radial equation can be written in terms of Bessel functions; more precisely, as a linear combination of the functions
\begin{equation}
u^{-\frac{1}{2}}J_{2q}(\Omega u^{\frac{1}{2}})\qquad \text{and}\qquad u^{-\frac{1}{2}}J_{-2q}(\Omega u^{\frac{1}{2}}) \, ,\label{Besssss}
\end{equation}
where $J_{2q}$ are Bessel functions. Therefore, up to confluent points, the general solution in the far region takes the form
\begin{equation}\label{Cien}
R_F(u)= A u^{-\frac{1}{2}}J_{2q}(\Omega u^{\frac{1}{2}})+B u^{-\frac{1}{2}}J_{-2q}(\Omega u^{\frac{1}{2}}) 
\end{equation}
where the subindex $F$ refers to the solution that is valid in the far region. For $u \gg 1$, the Bessel functions behaves like
\begin{equation}
    J_{2q}(\Omega u^{\frac{1}{2}})\simeq \sqrt{\frac{2}{\Omega \pi}}u^{-\frac{1}{4}}\off{\cos{\of{\Omega u^{\frac{1}{2}}-\frac{\pi}{2}\of{2q+\frac{1}{2}}}}+\mathcal{O}\of{{u^{-\frac{1}{2}}}}}
\end{equation}
and, then, the solution goes like
\begin{align}
    R_F(u) & \simeq \sqrt{\frac{2}{\Omega \pi}}u^{-\frac{3}{4}} \Big[ e^{i\Omega u^{\frac{1}{2}}}\of{A e^{-i\frac{\pi}{2}(2q +\frac{1}{2})}+B e^{-i\frac{\pi}{2}(-2q +\frac{1}{2})}} \nonumber \\
    & \qquad + e^{-i\Omega u^{\frac{1}{2}}}\of{A e^{i\frac{\pi}{2}(2q +\frac{1}{2})}+B e^{i\frac{\pi}{2}(-2q +\frac{1}{2})}}+ ... \Big]\, ; 
\end{align}
the ellipsis stand for subleading terms in powers of $1/u$. 

Demanding no incoming flux from past null infinity, for $\lambda=0$ we find
\begin{equation}
\frac{A}{B}=-e^{2\pi i q}=
-e^{i\pi \sqrt{1+K_{\ell} -(k_1+k_2)^2}}=e^{2\pi i \Delta_+}\, . \label{Simplon}
\end{equation}

It is worth noticing that, unexpectedly, the solution (\ref{Cien}) we find in the 5-dimensional case is substantially simpler than its 4-dimensional analog, cf. (3.50) in \cite{Porfyriadis:2014fja}. The fact that the solution to the equation above admits to be expressed in terms of Bessel functions --in contrast to the hypergeometric functions appearing in the 4-dimensional Kerr/CFT calculation-- leads to the rather simple expression (\ref{Simplon}), much simpler than the 4-dimensional expressions (3.50) and (4.37) in reference \cite{Porfyriadis:2014fja}. This is relevant because these expressions are what ultimately leads to the CFT$_2$ interpretation of a reflection coefficient. This phenomenon had already been observed in \cite{Bredberg:2009pv}; see footnote 9 in page 10 therein. The authors of \cite{Bredberg:2009pv} noticed that, unlike in 4 dimensions, where one encounters hypergeometric functions, in 5-dimensions the solution in the far region involves Bessel functions. This was interpreted as an indication that there may be
some kind of $SL(2, \mathbb{R})$ or even conformal symmetry associated with the far region in 4 dimensions: in the 4-dimensional case, the expression analogous to (\ref{Simplon}) has additional $\Gamma $-functions, exhibiting the characteristic form of a CFT$_2$ correlator. 

To implement the matching condition, we have to compare the expressions of the solution $R_F$ for $u \ll 1$ with the solution in the NHEMP geometry for $u\gg 1$. The former behaves like
\begin{equation}
    \label{eq:RF_small_u}
    R_F(u)\simeq -\of{\frac{\Omega}{2}}^{2q} \frac{B\, e^{2\pi i q}}{\Gamma(1+2q)}u^{-\frac{1}{2}+q}+\of{\frac{\Omega}{2}}^{-2q} \frac{B}{\Gamma(1-2q)}u^{-\frac{1}{2}-q} + ...
\end{equation}
On the other hand, the NHEMP solution
\begin{equation}
    R_{N}(u)=P \, W_{ip,\mu}\of{-\frac{i2\omega}{u}} + Q \, M_{ip,\mu}\of{-\frac{i2\omega}{u}}\, ,
    \label{eq:general_solution_NHEMP}
\end{equation}
for $u\gg 1$ behaves like
\begin{equation}\label{eq:RN_large_u}
  R_{N}(u)\simeq P c_1 u^{-\frac{1}{2}+\mu}+\of{Q(-2i\omega)^{\frac{1}{2}+\mu}+P c_2}u^{-\frac{1}{2}-\mu},
\end{equation}
with 
\begin{equation}
c_1=(-2i\omega)^{\frac{1}{2}-\mu}\frac{\Gamma(2\mu)}{\Gamma(\frac{1}{2} + \mu - ip)} \ \ \ \text{and} \ \ \ c_2=(-2i\omega)^{\frac{1}{2} + \mu}\frac{\Gamma(-2\mu)}{\Gamma(\frac{1}{2} - \mu - ip)}.
\end{equation}
Comparing \eqref{eq:RF_small_u} with \eqref{eq:RN_large_u} we can express $P$ and $Q$ as functions of $B$. This can be achieved if and only if $\mu=q$, a condition that is guaranteed by the identity $C=\gamma$ in the limit $\lambda \to 0$. In this way, we obtain
\begin{equation}
    B = Q (-2i\omega)^{\frac{1}{2}+q}\of{\frac{\Omega}{2}}^{2q}\Gamma(1-2q)\off{1-\of{\frac{\Omega}{2}}^{4q}e^{i2q\pi}(-2i\omega)^{2q}\frac{\Gamma(1-2q)^2}{\Gamma(1+2q)^2}\frac{\Gamma(\frac{1}{2} + q - ip)}{\Gamma(\frac{1}{2} - q - ip)}}^{-1}\label{Faltancosasche}
\end{equation}
together with
\begin{equation}
    P=-\of{\frac{\Omega}{2}}^{2q}\frac{e^{2\pi iq} \, B}{\Gamma(1+2q)}\, .
\end{equation}
Expression (\ref{Faltancosasche}) is different from its 4-dimensional analog; cf. equation (3.53) in \cite{Porfyriadis:2014fja}. More precisely, the 5-dimensional expression lacks a quotient of $\Gamma$-functions to admit the same CFT$_2$ interpretation than its 4-dimensional counterpart. The reason of this qualitative difference can be traced back to the Bessel functions in (\ref{Besssss}), which in 4 dimensions get rePlaced by hypergeometric functions. 

\subsection{Horizon boundary conditions}

So far, we have examined solutions in both the far and the near horizon regions. Now, let us impose conditions at the horizon: by imposing incoming boundary conditions for the modes on the horizon and outgoing boundary conditions in the far region, we will obtain a constraint for the wave numbers $k_{1,2}$ and the frequency $\tilde{\omega }$. The condition for purely outgoing flux at infinity was addressed above; the incoming boundary conditions at the horizon can be implemented by expanding the NHEMP solution \eqref{eq:general_solution_NHEMP} for $u \ll 1$, namely
\begin{equation}
    \begin{aligned}
        R_N(u) & \approx (-i2\omega)^{ip}Pe^{if(u)} + Q \Bigg[ (-i2\omega)^{-ip}e^{-if(u)}\frac{\Gamma(1 + 2\mu)}{\Gamma(\frac{1}{2} + \mu - ip)} \\
        & \qquad \qquad \qquad \qquad + (-1)^{\frac{1}{2} + \mu + ip}(-2i \omega)^{ip}e^{if(u)}\frac{\Gamma(1 + 2\mu)}{\Gamma(\frac{1}{2} + \mu + ip)} \Bigg] + ... \, \label{GGGG}
    \end{aligned}
\end{equation}
where we define the 
function
\begin{equation}\label{laf}
    f(u)=p\log{\frac{1}{u}}+\frac{\omega}{u} \, .
\end{equation}
We observe from this that both incoming and outgoing modes are present in this solution. Considering that near $u=0$, the $u^{-1}$ term dominates in \eqref{laf}, which is a decreasing dependence in the radial coordinate, the outgoing part of the solution (remember that $\Phi \sim R\of{u} e^{-i \omega t}$) is given by the first term inside the brackets in (\ref{GGGG}), namely
\begin{equation}
 Qe^{-if(u)}(-i2\omega)^{-ip}\frac{\Gamma(1+2\mu)}{\Gamma(\frac{1}{2}+\mu-ip)}\label{La111},
\end{equation}
which, therefore, must vanish. This can be achieved by either setting the coefficient $Q$ equal to zero or by studying the poles of the $\Gamma$-function in the denominator. Let us focus on the latter possibility: For $Q\neq 0$, (\ref{La111}) exhibit zeros at
\begin{equation}
\frac{1}{2}+\mu-ip=-n\quad\textrm{with } n\in\mathbb{Z}_{\geq 0}.
\end{equation}
With the definitions of $\mu$ in \eqref{lambda_y_mu}, of $p$ in \eqref{eq:whittaker_coeff_ilambda}, and of $\Delta_{\pm }$ in (\ref{eq:delta_definition}), we have
\begin{equation}
\begin{split}
\Delta_{\pm}-i\frac{1}{2}\of{\sqrt{\frac{a}{b}}k_1+\sqrt{\frac{b}{a}}k_2}&=-n
\end{split}
\label{eq:Q_equal_zero_condition}
\end{equation}
In order to satisfy these equations, and for $\text{sign} (a)=\text{sign}(b)$, we need  $\Delta_{\pm}$ to be real. This implies $b{k_1}+a{k_2}=0$, which makes \eqref{eq:Q_equal_zero_condition} become simply $\Delta_{\pm } = -n$, or equivalently
\begin{equation}\label{Delton}
    \begin{aligned}
        k_1 & = \frac{a}{a-b}\sqrt{1+K_{\ell}-(2n+1)^2} \\
        k_2 & = \frac{b}{b-a}\sqrt{1+K_{\ell}-(2n+1)^2}
    \end{aligned}
\end{equation}
This yields the quantization condition for the frequency $\tilde\omega$ in the limit $\lambda \to 0$, namely
\begin{equation}\label{La115}
    \tilde\omega =\frac{k_1+k_2}{a+b}=\frac{\sqrt{K_{\ell}-4n(n+1)}}{a+b} \, .
\end{equation}

Before concluding, let us briefly discuss the interpretation of the frequency (\ref{La115}). $\tilde \omega $ is the frequency of the solution in the asymptotic region that, according to (\ref{eq:omega-tilde}), matches the solutions in the near-horizon region. That is to say, all the frequencies of the solutions in the NHEMP geometry connect with (\ref{La115}) in the limit $\lambda \to 0$; this is due to the high redshift. Frequency (\ref{La115})  is the 5-dimensional analog to the superradiance threshold frequency of a Kerr black hole, cf. \cite{Frolov:2002xf}, i.e. the critical frequency to extract energy from the black hole by substracting angular momentum carried by the $k_{1,2}$ quantum numbers; this yields (\ref{eq:omega-tilde}). In addition, we observe that $\tilde \omega$ in (\ref{La115}) is the frequency for which the effective potential in the radial equation (\ref{eq:mp:radial_eq}) qualitatively changes, suppressing the terms that would be dominant for small $u$. In this sense, $\tilde \omega$ can be thought of as a penetration frequency. More precisely, in (\ref{eq:mp:radial_eq}) we can ask the functions $\alpha (\omega{(\tilde \omega)})$ and $\beta (\omega{(\tilde \omega)})$ to vanish, which precisely yields the first equality in (\ref{La115}). The second identity in (\ref{La115}) is nothing but the quantization condition for the frequency $\tilde \omega $ induced by the incoming flux condition in the horizon. A similar result is obtained for the near-extremal case.

\[\]

The authors thank J.J. Melgarejo for discussions on early stages of this collaboration. This work is partially supported by grants PIP-(2017)-1109, PICT-(2019)-00303, PIP-(2022)-11220210100685CO, PIP-(2022)-11220210100225CO, PICT-(2021)-GRFTI-00644.

\bibliography{references}

\begin{thebibliography}{10}

\bibitem{Guica:2008mu}
Monica Guica, Thomas Hartman, Wei Song, and Andrew Strominger.
\newblock {The Kerr/CFT Correspondence}.
\newblock {\em Phys. Rev. D}, 80:124008, 2009.
\newblock \href {https://arxiv.org/abs/0809.4266} {\path{arXiv:0809.4266}},
  \href {https://doi.org/10.1103/PhysRevD.80.124008}
  {\path{doi:10.1103/PhysRevD.80.124008}}.

\bibitem{Bardeen:1999px}
James~M. Bardeen and Gary~T. Horowitz.
\newblock {The Extreme Kerr throat geometry: A Vacuum analog of AdS(2) x S**2}.
\newblock {\em Phys. Rev. D}, 60:104030, 1999.
\newblock \href {https://arxiv.org/abs/hep-th/9905099}
  {\path{arXiv:hep-th/9905099}}, \href
  {https://doi.org/10.1103/PhysRevD.60.104030}
  {\path{doi:10.1103/PhysRevD.60.104030}}.

\bibitem{Bengtsson:2005zj}
Ingemar Bengtsson and Patrik Sandin.
\newblock {Anti de Sitter space, squashed and stretched}.
\newblock {\em Class. Quant. Grav.}, 23:971--986, 2006.
\newblock \href {https://arxiv.org/abs/gr-qc/0509076}
  {\path{arXiv:gr-qc/0509076}}, \href
  {https://doi.org/10.1088/0264-9381/23/3/022}
  {\path{doi:10.1088/0264-9381/23/3/022}}.

\bibitem{Anninos:2008fx}
Dionysios Anninos, Wei Li, Megha Padi, Wei Song, and Andrew Strominger.
\newblock {Warped AdS(3) Black Holes}.
\newblock {\em JHEP}, 03:130, 2009.
\newblock \href {https://arxiv.org/abs/0807.3040} {\path{arXiv:0807.3040}},
  \href {https://doi.org/10.1088/1126-6708/2009/03/130}
  {\path{doi:10.1088/1126-6708/2009/03/130}}.

\bibitem{Lu:2008jk}
H.~Lu, Jianwei Mei, and C.~N. Pope.
\newblock {Kerr/CFT Correspondence in Diverse Dimensions}.
\newblock {\em JHEP}, 04:054, 2009.
\newblock \href {https://arxiv.org/abs/0811.2225} {\path{arXiv:0811.2225}},
  \href {https://doi.org/10.1088/1126-6708/2009/04/054}
  {\path{doi:10.1088/1126-6708/2009/04/054}}.

\bibitem{Compere:2012jk}
Geoffrey Comp\`ere.
\newblock {The Kerr/CFT correspondence and its extensions}.
\newblock {\em Living Rev. Rel.}, 15:11, 2012.
\newblock \href {https://arxiv.org/abs/1203.3561} {\path{arXiv:1203.3561}},
  \href {https://doi.org/10.1007/s41114-017-0003-2}
  {\path{doi:10.1007/s41114-017-0003-2}}.

\bibitem{Castro:2010fd}
Alejandra Castro, Alexander Maloney, and Andrew Strominger.
\newblock {Hidden Conformal Symmetry of the Kerr Black Hole}.
\newblock {\em Phys. Rev. D}, 82:024008, 2010.
\newblock \href {https://arxiv.org/abs/1004.0996} {\path{arXiv:1004.0996}},
  \href {https://doi.org/10.1103/PhysRevD.82.024008}
  {\path{doi:10.1103/PhysRevD.82.024008}}.

\bibitem{Porfyriadis:2014fja}
Achilleas~P. Porfyriadis and Andrew Strominger.
\newblock {Gravity waves from the Kerr/CFT correspondence}.
\newblock {\em Phys. Rev. D}, 90(4):044038, 2014.
\newblock \href {https://arxiv.org/abs/1401.3746} {\path{arXiv:1401.3746}},
  \href {https://doi.org/10.1103/PhysRevD.90.044038}
  {\path{doi:10.1103/PhysRevD.90.044038}}.

\bibitem{Hadar:2014dpa}
Shahar Hadar, Achilleas~P. Porfyriadis, and Andrew Strominger.
\newblock {Gravity Waves from Extreme-Mass-Ratio Plunges into Kerr Black
  Holes}.
\newblock {\em Phys. Rev. D}, 90(6):064045, 2014.
\newblock \href {https://arxiv.org/abs/1403.2797} {\path{arXiv:1403.2797}},
  \href {https://doi.org/10.1103/PhysRevD.90.064045}
  {\path{doi:10.1103/PhysRevD.90.064045}}.

\bibitem{Bredberg:2009pv}
Irene Bredberg, Thomas Hartman, Wei Song, and Andrew Strominger.
\newblock {Black Hole Superradiance From Kerr/CFT}.
\newblock {\em JHEP}, 04:019, 2010.
\newblock \href {https://arxiv.org/abs/0907.3477} {\path{arXiv:0907.3477}},
  \href {https://doi.org/10.1007/JHEP04(2010)019}
  {\path{doi:10.1007/JHEP04(2010)019}}.

\bibitem{Myers:1986un}
Robert~C. Myers and M.~J. Perry.
\newblock {Black Holes in Higher Dimensional Space-Times}.
\newblock {\em Annals Phys.}, 172:304, 1986.
\newblock \href {https://doi.org/10.1016/0003-4916(86)90186-7}
  {\path{doi:10.1016/0003-4916(86)90186-7}}.

\bibitem{ronveaux1995heun}
A.~Ronveaux and F.M. Arscott.
\newblock {\em Heun's Differential Equations}.
\newblock Oxford science publications. Oxford University Press, 1995.
\newblock URL: \url{https://books.google.com.ar/books?id=5p65FD8caCgC}.

\bibitem{abramowitz1970handbook}
M.~Abramowitz and I.A. Stegun.
\newblock {\em Handbook of Mathematical Functions with Formulas, Graphs, and
  Mathematical Tables}.
\newblock Number v. 55,n.{\textordmasculine} 1972 in Applied mathematics
  series. U.S. Government Printing Office, 1970.
\newblock URL: \url{https://books.google.com.ar/books?id=ZboM5tOFWtsC}.

\bibitem{whittaker_watson_1996}
E.~T. Whittaker and G.~N. Watson.
\newblock {\em A Course of Modern Analysis}.
\newblock Cambridge Mathematical Library. Cambridge University Press, 4
  edition, 1996.
\newblock \href {https://doi.org/10.1017/CBO9780511608759}
  {\path{doi:10.1017/CBO9780511608759}}.

\bibitem{Frolov:2002xf}
Valeri~P. Frolov and Dejan Stojkovic.
\newblock {Quantum radiation from a five-dimensional rotating black hole}.
\newblock {\em Phys. Rev. D}, 67:084004, 2003.
\newblock \href {https://arxiv.org/abs/gr-qc/0211055}
  {\path{arXiv:gr-qc/0211055}}, \href
  {https://doi.org/10.1103/PhysRevD.67.084004}
  {\path{doi:10.1103/PhysRevD.67.084004}}.

\end{thebibliography}
\bibliographystyle{unsrturl}

\end{document}